\documentstyle[aps,prl,preprint,floats,epsfig]{revtex}

\providecommand{\cleoii}{\mbox{CLEO II}}
\providecommand{\cleoiiv}{\mbox{CLEO II.V}} 

\begin{document}

\preprint{\tighten\vbox{\hbox{\hfil CLNS 00-1668}
                        \hbox{\hfil CLEO 00-6}
}}

\title{Precise Measurement of $B^0-\overline{B^0}$ Mixing Parameters at the $\Upsilon(4S)$}  

\author{CLEO Collaboration}
\date{\today}

\maketitle
\tighten

\begin{abstract} 
We describe a measurement of $B^0-\overline{B^0}$ mixing parameters
exploiting a method of partial reconstruction of 
the decay chains $\overline{B} \rightarrow D^{*\pm}\pi^{\mp}$ and 
$\overline{B} \rightarrow D^{*\pm}\rho^{\mp}$.
Using 9.6 $\times 10^6 B\overline{B}$ pairs collected at the
Cornell Electron Storage Ring, we find $\chi_d = 0.198 \pm 0.013 \pm 0.014$,
$|y_d|<0.41$ at 95\% confidence
level, and $|\Re e(\epsilon_B)|<0.034$ at 95\% confidence level.
\end{abstract}
\newpage

{
\renewcommand{\thefootnote}{\fnsymbol{footnote}}

\begin{center}
B.~H.~Behrens,$^{1}$ W.~T.~Ford,$^{1}$ A.~Gritsan,$^{1}$
J.~Roy,$^{1}$ J.~G.~Smith,$^{1}$
J.~P.~Alexander,$^{2}$ R.~Baker,$^{2}$ C.~Bebek,$^{2}$
B.~E.~Berger,$^{2}$ K.~Berkelman,$^{2}$
% ...
\end{center}
 
\small
\begin{center}
$^{1}${University of Colorado, Boulder, Colorado 80309-0390}\\
$^{2}${Cornell University, Ithaca, New York 14853}\\
% ...
\end{center}

\setcounter{footnote}{0}
}
\newpage


The discovery of $B^0-\overline{B^0}$ mixing
in 1987 \cite{argus_first} signaled a large top
quark mass and allows the anticipation
of observable CP violating asymmetries in the
$B^0$ meson in the near future. Well-known values of the parameters
describing mixing will be necessary to extract precise values of
CP violating parameters, as planned at asymmetric $B$ factories 
\cite{babar_tdr}.  In addition, the size of 
$B^0-\overline{B^0}$ mixing is characterized
by the mass difference parameter $\Delta m_d$ and is
proportional to the square of
$|V_{tb}^*V_{td}|$, the magnitude of one side of the
``Unitarity Triangle'' \cite{babar_book} 
which describes some of the mathematical
constraints imposed by unitarity upon the elements of
the CKM matrix \cite{ckm}.  Accurate measurements of $\Delta m$ for
$B^0$ and $B^0_s$ mesons therefore provide an independent check on
our understanding of CP violation in the Standard Model.

	Mixing in the $B^0-\overline{B^0}$ system may be described by
the parameters  $x_d$, $y_d$, $p$, and $q$.
The parameter $x_d=\Delta m_d$/$\Gamma_d$, where $\Delta m_d$ is
the mass difference between the heavy and light 
eigenstates $B_H$ and $B_L$ and $\Gamma_d$ is the average
natural decay width.
 Similarly, $y_d$ is the normalized lifetime difference
between the two eigenstates, and 
can be written as
$\Delta \Gamma_d$/$2 \Gamma_d$.
The parameters $p$ and $q$ describe the
$B^0$ and $\overline{B^0}$ amplitudes, respectively, 
in the eigenstates $B_H$ and $B_L$.
When the $\Upsilon(4S)$ is produced
with symmetric energy electron positron 
collisions, the experimentally accessible quantity is $\chi_d$, 
the time-integrated probability for an
initially produced $B^0$ or $\overline{B^0}$ meson to decay as
its CP conjugate.
It may be written in terms of $x_d$ and $y_d$:

\begin{center}
\begin{eqnarray}
\chi_d = \frac{\Gamma(B^0 \rightarrow \overline{B^0})}
{\Gamma(B^0 \rightarrow B^0) + \Gamma(B^0 \rightarrow \overline{B^0})}
\simeq \frac{x_d^2 + y_d^2}{2(1+x_d^2)}.
\label{chi_xy}
\end{eqnarray}
\end{center}

\noindent Under certain assumptions, a measurement of
$\chi_d$ can be combined with direct determinations
of $\Delta m_d$ and the $B$ meson lifetime in order
to extract $y_d$.  The ancillary variable, $\epsilon_B$, is
analogous to the $K^0$-mixing parameter $\epsilon$, and is
defined through the relation 
$p = ({1+\epsilon_B}$)/${\sqrt{2(1+|\epsilon_B|^2)}}$.
Limits on $\epsilon_B$ can be extracted by searching for a
CP violating asymmetry in the events where mixing has occurred.
The mixing parameters $y_d$ and $\epsilon_B$ are
both expected to be of order $10^{-2}$ \cite{babar_book}
with considerable uncertainty.

	In this letter, we report new measurements of 
the $B^0-\overline{B^0}$ mixing
parameters measured at the  $\Upsilon(4S)$ resonance.
We attempt to determine the beauty quantum
number (or flavor) at decay for both of the $B$ mesons produced
in the $\Upsilon(4S)$ decay  using a
novel method
subject to systematic uncertainties very different from previous 
measurements \cite{argus_first},\cite{cleo_mix}.
When the decay flavors of the two $B$ mesons in the event coincide, 
it indicates that the second $B$ has undergone 
mixing in the interval between the decays of the two
$B$ mesons. The flavor of one $B$ meson at decay is
tagged by a high-momentum lepton originating from 
the decay chain $B \rightarrow X \ell \nu$.
The flavor at decay of the remaining $B$
meson is determined through 
partial reconstruction of the 
decay chain $\overline{B^0} \rightarrow D^{*+}h_W^{-}$ (charge
conjugate modes implied), 
where $h_W^-$ refers either to a $\pi^-$ or $\rho^-$.
The electric charge of the $h_W$ identifies (tags) 
the value of the $B$ flavor at the
time of its decay. (We assume the double Cabibbo suppressed decay  
$\overline{B^0} \rightarrow D^{*-}h_W^{+}$ is negligible.)
By employing the hadronic flavor tag for one $B^0$ in the
event, this method sacrifices statistical accuracy
relative to methods where a semileptonic decay is
used to tag the flavor of both $B$ mesons in the event.
However, the systematic error due to the uncertainty in the
charged to neutral $B$ meson production ratio
at the $\Upsilon(4S)$ that
dominates measurements of $\chi_d$ at the
$\Upsilon(4S)$ using dileptons is substantially reduced \cite{pdg}.
As a result this method results in a significant improvement
in precision over previous measurements of $\chi_d$ at the 
$\Upsilon(4S)$.

	Four charge combinations of hadrons and
leptons are possible: $h_W^+\ell^+$, $h_W^-\ell^-$, 
$h_W^+\ell^-$, and $h_W^-\ell^+$.  
In the absence of backgrounds or mistags, these
correspond to the four
flavor combinations $B^0 B^0$, $\overline{B^0} \, \overline{B^0}$,
$B^0 \overline{B^0}$, and  $\overline{B^0} B^0$, respectively.
Then,

\begin{center}
\begin{eqnarray}
\chi_{d} = \frac{h_W^+ \ell^+ + h_W^-\ell^-}{h_W^+ \ell^+ + h_W^-\ell^- + h_W^+ \ell^- + h_W^-\ell^+}.
\label{chi_raw_def}
\end{eqnarray}
\end{center}

\noindent In practice, the raw counts recorded in each of the combinations must be
corrected for processes that incorrectly tag the $B$ decay flavor (mistags).
Mistags may be due either to leptons not arising from the primary
decay $B \rightarrow X \ell \nu$ or hadrons not arising from the
hypothesized decay chain $\overline{B^0} \rightarrow D^{*+}h_W^-$.
Backgrounds (non-$\overline{B^0}$ events), which can contribute either to 
the denominator or numerator, must be
subtracted.

	\par The data were recorded at the Cornell Electron Storage Ring (CESR)
with two configurations of the
CLEO\ detector called {\cleoii}~\cite{kubota} and {\cleoiiv}.
In  the  \cleoiiv\ configuration, the innermost wire chamber was 
replaced with a
precision three-layer silicon vertex detector (SVX)~\cite{svx}.
The results presented here are based upon an integrated luminosity of 
9.1 $\rm fb^{-1}$ of $e^+e^-$ data taken at the $\Upsilon(4S)$ energy 
and 4.4 $\rm fb^{-1}$ taken an average of 60 MeV below $B \overline B$ threshold. 
The Monte Carlo simulation of the CLEO detector was
 based upon 
GEANT~\cite{GEANT} and simulated events
were processed in the same manner as the data. 

	The method of partial reconstruction used in this Letter
has been described in detail elsewhere \cite{gronberg}.
By observing the $h_W$
and the soft pion from the decay $D^{*+} \rightarrow D^0 \pi_s^+$, 
we deduce the kinematics
of the decay chain $\overline{B^0} \rightarrow D^{*+} h_W^-$ 
without reconstruction of the $D^0$.  
We denote the charged pion from the $h_W$ as $\pi_f$.

\begin{figure}[ht]
   \centering \leavevmode
   \epsfxsize=6 cm
\epsfbox{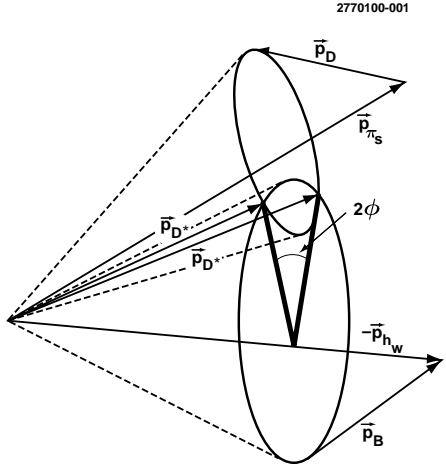}
   \caption{The definition of the angle $\phi$.
The lower cone shows constraints of the $B$ decay kinematics, and the
upper cone shows those of the $D^*$ decay kinematics;
their intersection defines the two solutions for the $D^*$ momentum.}
   \label{phi_fig}
\end{figure}

The hadronic $B^0$ decay may be described by three angles.  
The angle formed, in the $B$ ($D^*$) rest frame, between the 
$D^* (D^0)$ flight direction and the direction of the lab
frame, is called $\theta_B^* (\theta_{D^*}^*)$.
A larger value of $\cos \theta_B^* (\cos \theta_{D^*}^*) $
corresponds to a higher momentum of the $h_W$($\pi_s$).
The third angle, $\phi$, is the angle 
between the plane of the  $\overline{B^0} \rightarrow D^{*+} h_W^-$ 
decay and
the plane that contains the $h_W^-$ and the $\pi_s^+$, 
as shown in Fig. \ref{phi_fig}.  All three angles have distinctive
distributions for signal and background.  The $\cos\theta_B^*$ 
distribution is constant 
for signal because the $B$ meson is a scalar particle.  The
distribution of $\cos\theta_{D^*}^*$
shows the 100\% polarization in the $D^{*+}\pi^-$ mode from conservation
of angular momentum,
and shows the 87\% polarization that has been measured in the $D^{*+}\rho^-$
mode.  The distribution of $\cos\phi$ is a combination of
the $\pi_s$ and $h_W$ momenta and the angle between them.  For most
signal events it reconstructs inside the physical region 
$-1 < \cos\phi < 1$.  For non-signal events as well as
signal events with imperfect measurements of the pion momenta, 
$\cos\phi$ can be calculated
but may fail to have a physical value.

	In order to select hadronic $D^{*+}h_W^-$ decay
candidates for the analysis, both the 
$\pi_s$ and $\pi_f$ candidate tracks must be well-reconstructed 
and consistent with originating at the $e^+e^-$ interaction
point, and must not be identified as a lepton.
We reconstruct $\rho^{\pm}$ candidates from  $\pi^{\pm}_f\pi^0$ combinations,
where the $\pi^0$ is formed from a pair of photon candidates.

The $h_W$ momentum is required to fall in the kinematically
allowed range for  $\overline{B^0} \rightarrow D^{*+} h_W^-$ decays, assuming
$E_{B^0} = E_{beam}$.
We require the momentum of the $\pi_s$ to be below 300 MeV/$c$.
The $\pi_s$ and $\pi_f$ are required to have opposite electrical charges.
We require $|\cos\phi|<7$ and  $\cos\theta_{h_W\pi_s} <  -0.8$, where $\theta_{h_W\pi_s}$ 
is the angle between the $h_W$ and $\pi_s$.
In 8\% of selected events, there is more than one combination of
charged tracks that satisfy these criteria.
In this case, we select the one that is reconstructed
as $D^*\pi$ rather than $D^* \rho$.  
If more than one combination still remains, we choose the one for which
the value of $\cos \phi$ is nearest 0.6 (the peak of the
signal distribution.)
The resulting bias in the $\cos\phi$ shape
has a negligible effect upon our measurement of the mixing parameters.
The events satisfying these criteria are used to
determine  the sample composition.
For the events we use to measure yields (and therefore the mixing parameters),
we use more restrictive criteria, requiring $\cos\theta_{h_W\pi_s} <-0.95$,
$|\cos\phi|<2$, and $|\cos\theta_{D^*}^*|<1$.

	Lepton candidates are selected by requiring
that the track is well-reconstructed,
consistent with originating from
the $e^+e^-$ interaction point,
and well identified as either an
electron or a muon.
We require that the momentum of the lepton candidate is 
greater than 1.4 GeV/$c$
and we use the angle between the lepton and the $\pi_s$ to suppress
semileptonic decays from the unreconstructed $D^0$.  We veto
leptons from the decays $J/\psi \rightarrow \ell^+ \ell^-$.
If the event has been partially reconstructed as $D^{*+} \pi^-$, we 
require that the lepton form a large angle with the thrust axis of the remainder of
the event
in order to suppress $e^+e^- \rightarrow q\overline{q}$ backgrounds, 
where $q=\{u,d,s,c\}$.
If more than one lepton candidate in an event satisfies the criteria,
we select the highest momentum candidate.

For the  events selected by these criteria,
the $B$ decay modes contributing to the $D^{*+}h_W^-$ 
candidates
can be divided into five categories:
signal ($D^{*+}\pi^-$ and $D^{*+}\rho^-$),
other two-body and semi-leptonic $\overline{B^0}$ decays 
(such as $\overline{B^0} \rightarrow D^{+} \rho^-$ and  $\overline{B^0} \rightarrow D^{+} 
\ell^- \overline{\nu}$), 
two-body and semi-leptonic 
$B^{\pm}$ decays (such as $B^{-} \rightarrow D^{*0}\pi^-$ and
$B^{-} \rightarrow D^{*0} \ell^- \overline{\nu}$),
random combinatoric backgrounds, and events of the type
$e^+e^- \rightarrow q\overline{q}$ (continuum).
For the distributions of two-body $B^0$ 
and $B^+$ decays in
$\cos \theta_{D^*}^*$ and $\cos\phi$, we rely on the simulation.
We include 10 two-body and semi-leptonic decay modes of the $B^0$ in the
definition of two-body $B^0$ decays, and 12 in the
definition of the two-body $B^+$ decays \cite{mythesis}.  
These  decays are well-measured \cite{pdg} and
are reliably modeled by the simulation.
Combinations of $h_W$ and $\pi_s$ that satisfy the analysis
requirements, yet originate from neither
the signal decays nor the two-body decays, are considered random 
combinatoric backgrounds.
To model these, we use a synthetic distribution in $\cos \theta_{D^*}^*$ 
and $\cos\phi$, generated by combining track pairs drawn at random from
the observed spectrum of all $B\overline{B}$ track momenta in data.
The cosine of the angle between them, $\cos\theta_{h_W\pi_s}$,
is distributed uniformly.
The simulation predicts that the distribution generated with this
procedure provides
an excellent approximation to the distributions of these decays,
and indicates that $40\%$ of the combinatoric background comes from
$B^+B^-$ events.
Distributions from continuum $q\overline{q}$ production
are directly measured in data taken below the $\Upsilon(4S)$ resonance.

	In order to determine the composition of our event sample, we
divide the sample into two subsets 
based on the value of $\cos\theta_{h_W\pi_s}$.
Events for which $-0.9 <\cos\theta_{h_W\pi_s}<-0.8$ are
the sideband sample; events for which
$\cos\theta_{h_W\pi_s}<-0.95$ make up the signal sample.
We perform a binned two-dimensional 
maximum likelihood fit \cite{barlow} simultaneously to the 
$\cos\phi$ and $\cos\theta_{D*}^*$ distributions of 
the signal sample  and the sideband sample.
The fit determines the normalization of the first
four categories of events; that of the continuum
is fixed by the relative luminosity of the data taken at and below
the $\Upsilon(4S)$.

\begin{figure}[ht] 
   \centering \leavevmode 
        \epsfxsize=8 cm
	\epsfbox{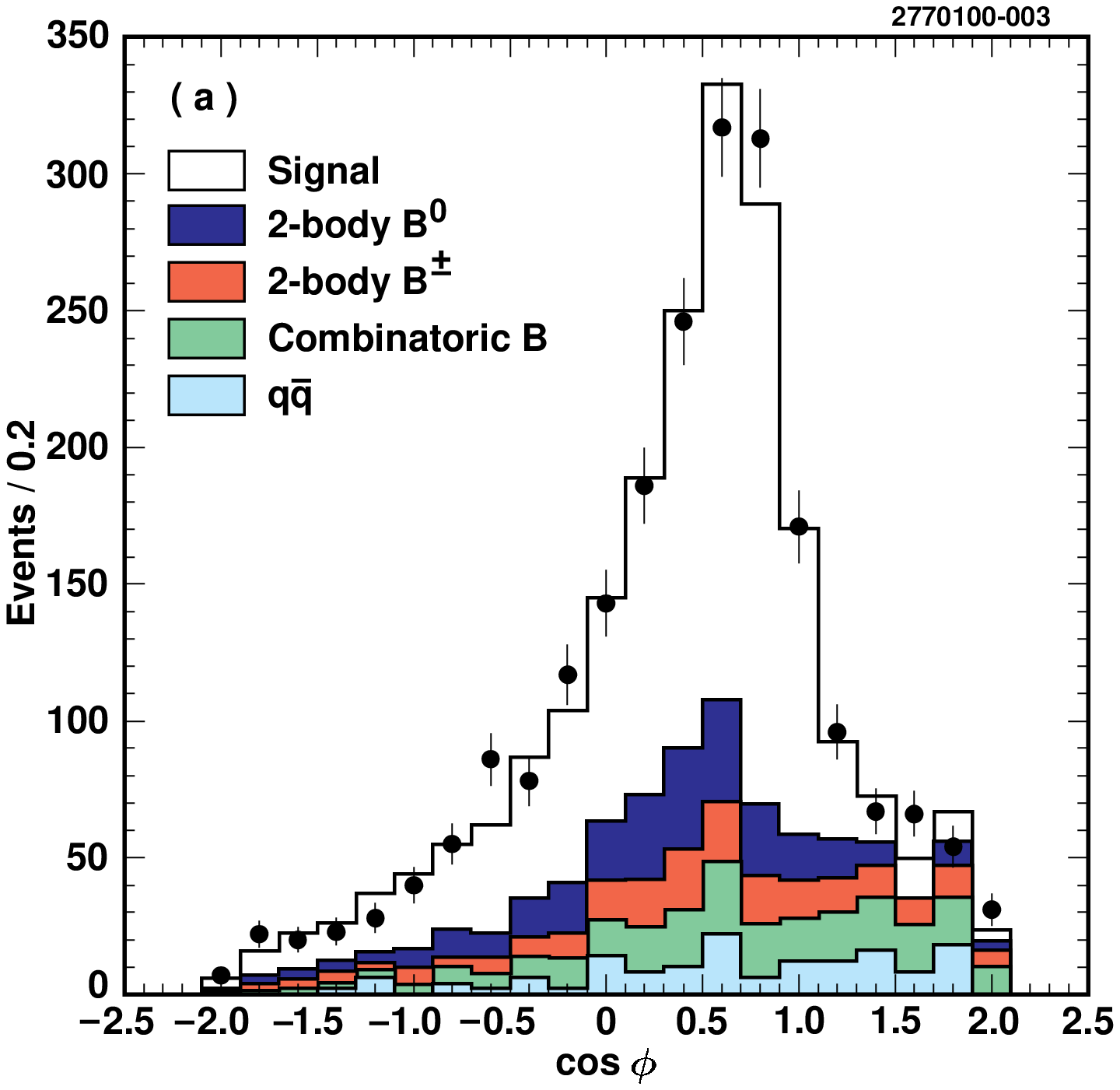}
        \epsfxsize=8 cm
	\epsfbox{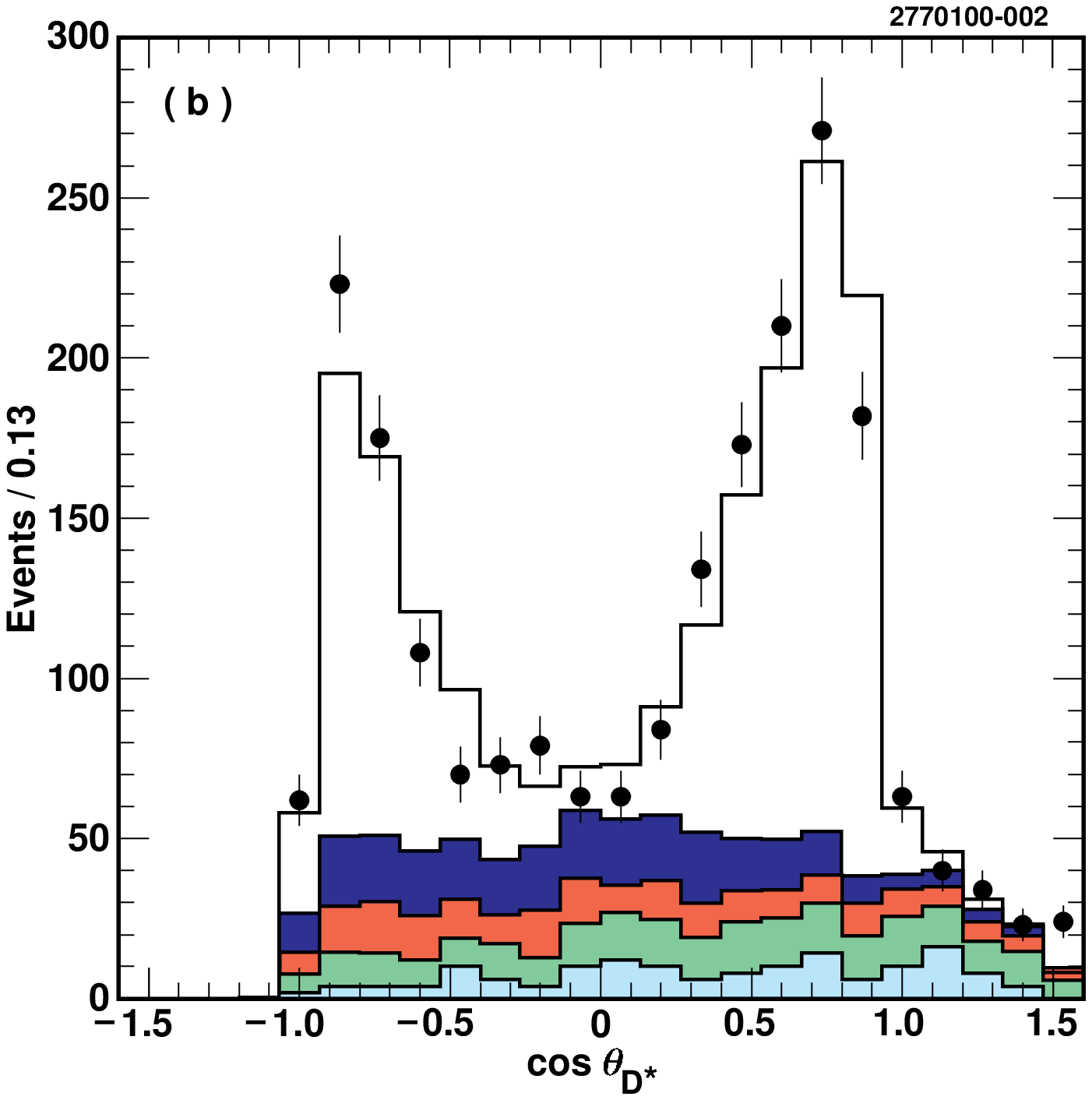}
\caption{Projections of the fit to the signal sample onto (a)
the $\cos\phi$ axis and (b) the
$\cos\theta_{D^*}$ axis.
In each plot, the points are the data and the histograms are the
best-fit shapes from the simulation.
} 
\label{proj_fit_sig}
\end{figure}

	The results of the fit are shown in Table \ref{absolute}. 
The two projections of the fits in the mixing sample (more
restrictive selection criteria) are
shown in Fig. \ref{proj_fit_sig}.  
We find that the fit has a
Baker-Cousins $\chi^2$ \cite{baker} of 129.7 for 151 degrees of freedom.
From simulations we find that this corresponds to a
confidence level of 91\%.

\begin{figure}[ht] 
   \centering \leavevmode 
        \epsfxsize=8 cm
	\epsfbox{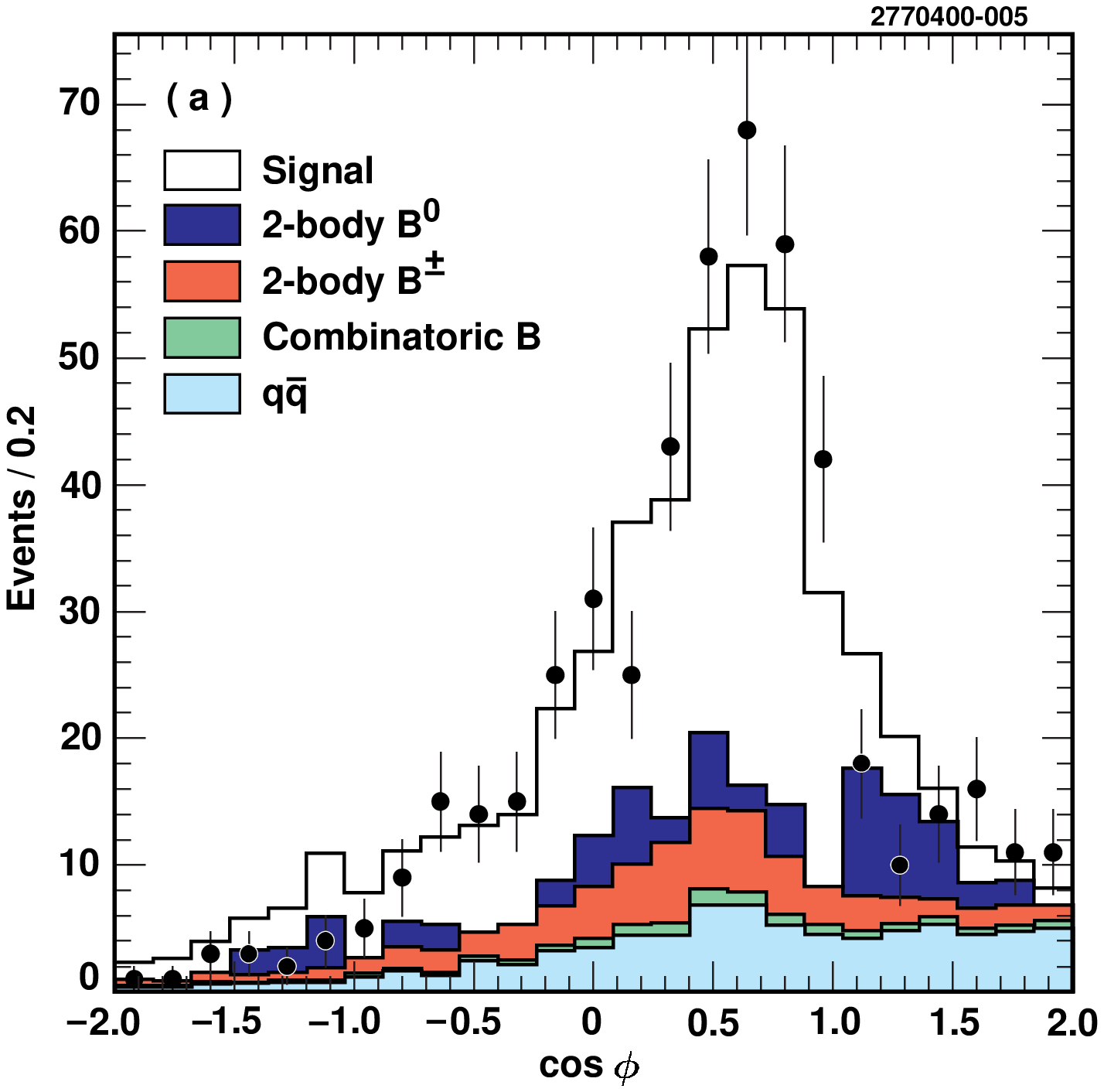}
        \epsfxsize=8 cm
	\epsfbox{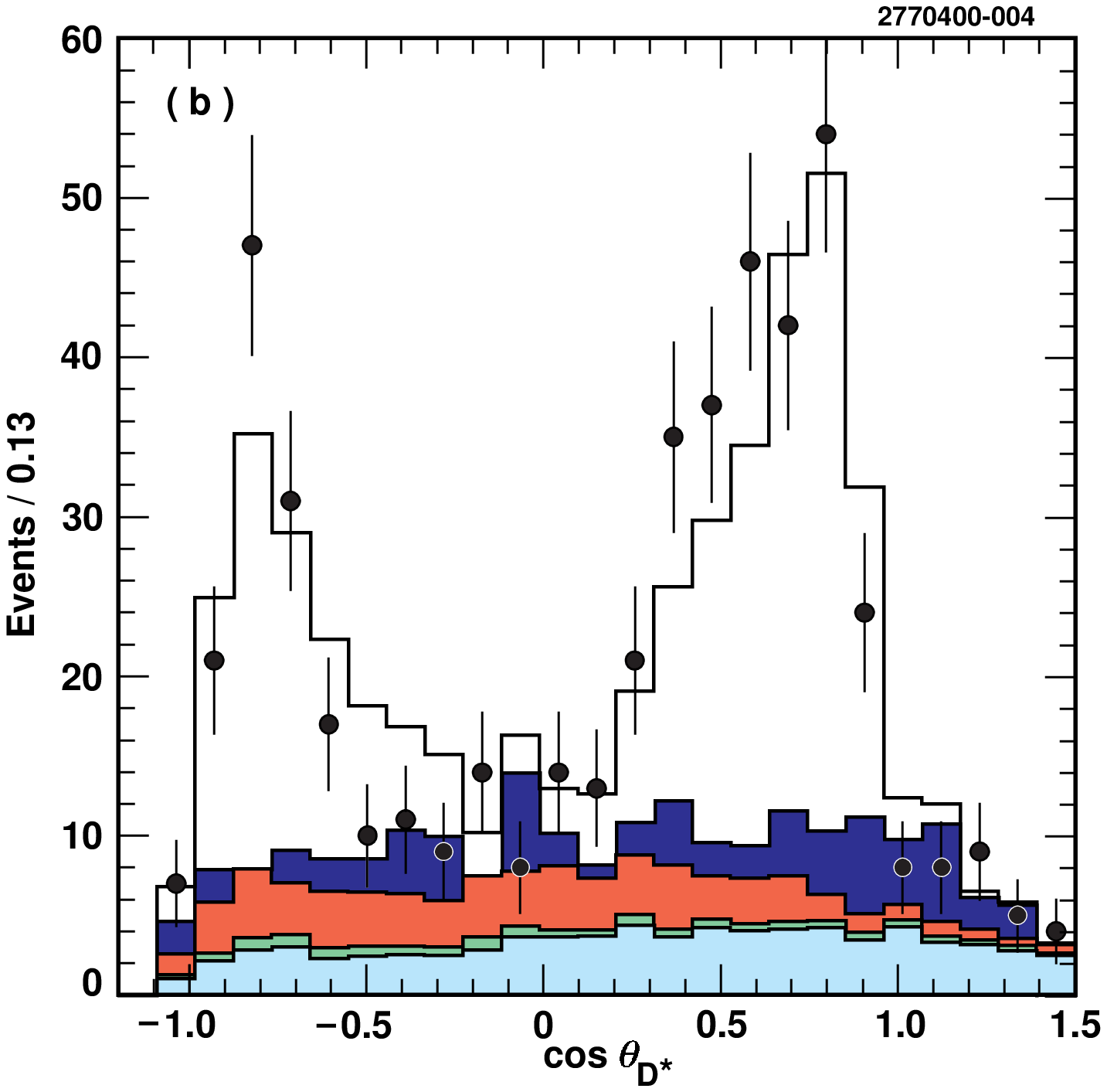}
\caption{Projections of the fit to the sample of mixed events onto (a)
the $\cos\phi$ axis and (b) the
$\cos\theta_{D^*}$ axis.
In each plot, the points are the data and the histograms are the
best-fit shapes from the simulation.
} 
\label{proj_fit_mixed}
\end{figure}

\noindent We also show in
Fig. \ref{proj_fit_mixed}  that the
distributions of the subset of events in Fig. \ref{proj_fit_sig}
that contribute to the numerator of Eqn. \ref{chi_raw_def}, ($h^{\pm}l^{\pm}$
events), are well described by the fit.

	 The partially reconstructed hadronic tag may
incorrectly identify the flavor of the decaying $B$ meson.
The dominant source of mistagged events is $B^0$ candidates
formed from random combinations of tracks in
$B^+B^-$ or $B^0\overline{B^0}$
events.
We determine that the mistag rate of combinatoric events 
is (21$\pm$12)\% (where the uncertainty
is statistical only),
using a separate sample of fully-reconstructed  $B\rightarrow D^* \ell \nu$ 
decays in the data.
We combine the composition as determined by the
fit (Table \ref{absolute}) with the mistag rates for each individual
component to determine the hadronic mistag rate, also shown in
Table \ref{absolute}. 
We calculate a total hadronic mistag rate of (3.1$\pm$1.2)\%.
The uncertainty includes the statistical uncertainties from the fit
and in the random combinatoric mistag rate.

\begin{table}
\centering
\caption{The composition of the mixing sample, as determined by the fit.
For comparison, the continuum subtracted data sample size is 1865 events. 
The mistag fraction uncertainties are statistical only.}
\begin{tabular}{ccc} 
Event Type      	& Fit                  	& Mistag Fraction\\ \hline
Signal          	& 1241$\pm$52         	& 0.0006$\pm$0.0006\\
Two-body $B^0$    	& 262$\pm$60         	& 0.020$\pm$0.005\\
Two-body $B^{\pm}$	& 172$\pm$45         	& 0.050$\pm$0.010\\
Combinatoric		& 192$\pm$21	     	& 0.21$\pm$0.12\\
Total             	& 1867$\pm$92          	& 0.031$\pm$0.012\\
\end{tabular}
\label{absolute}
\end{table}

	The lepton in the event may also
mistag the flavor of the decaying $B$ meson 
for several reasons.
Leptons may arise from the secondary decay chain $B \rightarrow DX, 
D \rightarrow X \ell \nu$.
The magnitude of this source is well-constrained by measurements of
the $B \rightarrow D X$ spectrum \cite{d_spectrum} and the known form 
factors governing $D$ semileptonic decays \cite{burchat}.
We also correct for leptons from $q\overline{q}$ events,
$B \rightarrow \psi X$, $\psi \rightarrow \ell^+\ell^-$ events \cite{pdg},
misidentified hadrons \cite{mythesis}, leptons from $D_s^+$\cite{ds_meas} and other upper vertex ($b \rightarrow \overline{c}$)
production \cite{flav_specific}, in-flight decays, $\pi^0$ Dalitz decays,
$\gamma$ conversions, and $\delta$ rays \cite{mythesis}.
Altogether we find that (3.6$\pm$0.5)\% of electrons and 
(3.8$\pm$0.5)\% of muons incorrectly tag the $B$ decay flavor.
The uncertainties are the total systematic uncertainties obtained
by adding in quadrature the uncertainties associated with
each of the input branching fractions, spectra, and fake rates.


	
	The yields for the mixing sample (more
restrictive selection criteria) in the possible charge
combinations, and subsequent corrections, 
are summarized in Table \ref{counts_corr}.
We correct the continuum subtracted
raw yields of Table \ref{counts_corr}
for the mistag levels that we have
determined for the leptonic and hadronic tags, 
then subtract $B^+B^-$ background, which
contributes to the denominator even when the beauty quantum number
is correctly reconstructed.  The total charged $B$ background is
(13.3$\pm2.5)\%$, which is the sum of the two-body $B^+B^-$ decays
and the $40\%$ of the combinatoric background
that is attributed to $B^+B^-$.
The fully-corrected result is $\chi_d = 0.198 \pm 0.013 \pm 0.014$.
The systematic uncertainties are listed in Table \ref{syst}.

\begin{table}
\centering
\caption{Data event yields, sequentially subjected to the corrections, and the
value of $\chi_d$ computer from these yields.  The uncertainties are statistical only.}
\begin{tabular}{ccccc}
Correction                    & $h^\pm\ell^\pm$& $h^{\pm}\ell^{\mp}$ & Corrected $\chi$\\ \hline
None (raw yield) 		& 458$\pm$21	& 1524$\pm$39	& 0.231$\pm$0.010 \\
Continuum			& 401$\pm$23	& 1464$\pm$40	& 0.215$\pm$0.011 \\
Electron mistags		& 377$\pm$23    & 1487$\pm$31   & 0.202$\pm$0.011  \\
Muon mistags			& 359$\pm$24	& 1505$\pm$42	& 0.193$\pm$0.012  \\
Hadron mistags			& 321$\pm$25	& 1543$\pm$43	& 0.172$\pm$0.012  \\
$B^+B^-$ Background		& 321$\pm$25	& 1294$\pm$38	& 0.198$\pm$0.013  \\
\end{tabular}
\label{counts_corr}
\end{table}


\begin{table}
\centering
\caption{Summary of systematic uncertainties for $\chi_d$.}
\begin{tabular}{cc} 
Source						& Uncertainty	\\ \hline
Hadronic Mistag Fraction    			& $0.009$     \\
$B^{\pm}$  Background             		& $0.007$    \\
Two-Body Distributions         			& $0.006$    \\
Mixed:Unmixed Efficiency Difference		& $0.004$   \\ 
Lepton Mistag Fraction          		& $0.003$     \\
Mixed:Unmixed Mistag Rate Difference		& $0.003$	 \\
Signal Shape					& $0.002$	 \\
Combinatoric Shape              		& $0.002$    \\ \hline
Total                           		& $0.014$    \\ 
\end{tabular}
\label{syst}
\end{table}

	The largest systematic uncertainty in $\chi_d$ 
is due to the uncertainty in the total hadronic mistag rate
which in turn is dominated by the uncertainty in
the mistag rate
of combinatoric background events.
The systematic uncertainty in the combinatoric background
mistag rate is determined by comparing the mistag rates
of energetic pions in data and simulated events, using samples
in which the $B$ flavor has been tagged using 
the decay $D^{*+}\ell^-\overline{\nu}$.  
Smaller contributions to the total mistag rate come from
the statistical uncertainty of the fit and uncertainty in
the two-body mistag rates.
We evaluate mistag rates for hadronic and leptonic
tags independently, assuming no correlation.
The difference in efficiencies and mistag rates for
mixed and unmixed events are both
found to be consistent with zero in
large samples of simulated signal events, and
the uncertainty in $\chi_d$ reflects the statistical uncertainty
of the finite simulation samples.
We assign a systematic uncertainty in $\chi_d$ due
to the uncertainty in the mistag rate totaling 9\%.

	The uncertainty in the $B^+B^-$ background is dominated by
uncertainty in the percentage of random combinations that are due 
to $B^+B^-$ decays.
The assigned uncertainty allows the fraction
of random combinations arising from $B^+B^-$ decays
to vary uniformly from 0 to 100\%.

	The uncertainty due to the distributions
in $\cos\theta^*_{D^*}$ and $\cos\phi$ of two-body 
decays is evaluated by repeating the analysis,
varying in turn each two-body decay mode's weight
according to the experimental uncertainty 
in its branching fraction \cite{pdg} and
the limited statistics of the simulation.

	The uncertainty due to the shape of the signal distribution
is assessed by examining the
variation of $\chi_d$ as the data are refit with
modified signal distributions.
Modifications include variations in $D^* \rho$ polarization,
overall tracking efficiency, solution multiplicity, and 
beam energy.
The uncertainty due to the fitting distribution of combinatoric decays
is evaluated by considering possible variations in the momentum
spectrum of random tracks.  
We consider three variations, corresponding to decreasing
the average momentum of soft tracks, increasing the
average momentum of high-momentum tracks 
and changing the shape of the low-momentum spectrum as if 
there were 50\% more $D^*$'s than expected.

	By comparing the yield of $B^0 B^0$ candidate events to
the yield of $\overline{B^0} \, \overline{B^0}$ candidates,
we can limit $\epsilon_B$ from this
measurement  \cite{dir_cpnote}. 
We compare $\chi_d$ in events 
with positively charged leptons to $\chi_d$ with negatively
charged leptons,
defining $\chi_\pm $= $h_W^\pm \ell^\pm$/($h_W^\mp \ell^\pm + h_W^\pm\ell^\pm$)
and $ A_{CP}$ = ($\chi_+ - \chi_-$)/($\chi_+ + \chi_-$),
where $A_{CP} = 4 \Re e(\epsilon_B)$.
Charge asymmetries in lepton identification
cancel with this method.
From studies of detection asymmetries for hadrons 
and from measurements of hadronic fake 
contributions, we find negligible systematic bias in the measurement and
estimate a systematic uncertainty of 1.4\% on $A_{CP}$.  
We determine $A_{CP}$ = 0.017$\pm$0.070$\pm$0.014,
corresponding to $\Re e(\epsilon_B)=0.004 \pm 0.018 \pm 0.003$,
or $|\Re e(\epsilon_B)|< 0.034$ at the
95\% confidence level.

	We are also 
able to provide a non-trivial limit on $y_d$ using Eqn.(\ref{chi_xy})     
under two assumptions.
We assume 
that any possible $\Delta \Gamma_d$ has negligible
impact upon the extraction of $\Delta m_d$ from the experimental results
listed in \cite{pdg}, and we assume that 
indirect CP violation is not present ($|p/q|=1$).
We combine our measurement of $\chi_d$ with the values $\Delta m_d = 0.464 \pm 0.018 \, {\rm ps^{-1}}$ and 
$\tau_{B^0} = 1.56 \pm 0.04 \, {\rm ps}$ \cite{pdg}, to find $|y_d| < 0.41$ at
95\% confidence level.

	We have described a measurement of 
$B^0-\overline{B^0}$ mixing parameters $\chi_d$ and $\epsilon_B$, 
and have combined our result with
direct measurements of $\Delta m_d$ to extract limits on 
$y_d$.
We exploit a method of partial reconstruction of 
the decay chains $B \rightarrow D^{*\pm}\pi^{\mp}$ and 
$B \rightarrow D^{*\pm}\rho^{\mp}$ subject to systematic
uncertainties different from previous measurements.
We note that this result is independent of previous CLEO mixing
analyses which used leptons to tag the beauty quantum numbers
at decay for both $B$ mesons in the event \cite{cleo_mix}.
This measurement provides the first non-trivial limits on $y_d$.

	We gratefully acknowledge the 
effort of the CESR staff in providing us with
excellent luminosity and running conditions.
This work was supported by 
the National Science Foundation,
the U.S. Department of Energy,
the Research Corporation,
the Natural Sciences and Engineering Research Council of Canada, 
the A.P. Sloan Foundation, 
the Swiss National Science Foundation, 
the Texas Advanced Research Program,
and the Alexander von Humboldt Stiftung.


\begin{thebibliography}{99}
\bibitem{dir_cpnote}
{Under the assumption that there is no direct $CP$ violation in the
decay $\overline{B^0} \rightarrow D^{*+} h_W^-$.}
\end{thebibliography}
\end{document}